\documentclass[twocolumn, prl]{revtex4}

\usepackage{graphicx}% Include figure files
\usepackage{dcolumn}% Align table columns on decimal point
\usepackage{bm}% bold math

\begin{document}

\title{ N\'eel and Valence-Bond Crystal phases of the Two-Dimensional Heisenberg Model on the Checkerboard Lattice}

\author {S. Moukouri }

\affiliation{ Department of Physics and  Michigan Center for 
          Theoretical Physics \\
         University of Michigan 2477 Randall Laboratory, Ann Arbor MI 48109}

\begin{abstract}
I use an improved version of the two-step density matrix renormalization
group method to study ground-state properties of the 2D Heisenberg model on the
checkerboard lattice. In this version the  Hamiltonian is projected on a tensor 
product of two-leg ladders instead of chains. This allows investigations of 
2D isotropic models. I show that this method can describe both the 
magnetically disordered and ordered phases. The ground-state phases of 
the checkerboard model
as $J_2$ increases are: (i) N\'eel with $Q=(\pi,\pi)$, (ii) a valence 
bond crystal (VBC) 
of plaquettes, (iii) N\'eel with $Q=(\pi/2,\pi)$, and (iv) a VBC of crossed
dimers. In agreement with previous results, I find that at the isotropic 
point $J_2=J_1$, the ground state is made of weakly
interacting plaquettes with a large gap $\Delta \approx 0.67 J_1$ to triplet 
excitations.

\end{abstract}

\maketitle

Frustration-induced magnetically disordered phases in two dimensions (2D)
recently have attracted substantial interest \cite{lhuillier}. 
 Frustrated magnets are known to display unconventional ground states 
with, in some cases, a large set of low-lying degenerate singlet excitations 
that are still not well understood. Among models of frustrated systems, the 
Heisenberg model on the checkerboard lattice (HMCL) has recently been 
intensively studied by various techniques \cite{lieb,palmer,fouet,
canals,berg,sfb}. This model is seen as 
a first step in the investigation of the 3D pyrochlore model. 
 The emerging picture is that at the isotropic point
($J_1=J_2$), the HMCL spontaneously breaks the lattice's translational 
symmetry. The ground state is a singlet made of a collection of weakly coupled 
plaquettes with a large gap, $\Delta \approx 0.7 J_1$, to triplet excitations.
Away from the isotropic point, the situation is less clear.  There
is no single method which can capture the full phase diagram.

In this letter, I introduce an improved version of the two-step density-matrix 
renormalization group (TSDMRG) \cite{moukouri-TSDMRG, moukouri-TSDMRG2} which, 
as I will
show, is very convenient in the study of the HMCL and other 2D frustrated 
models. This new version is based on  using the two-leg ladder, instead of chains,
 as the starting point to build the 2D lattice.
 The main insight in using the two-leg ladder to construct the
2D lattice comes from large $N$ predictions 
\cite{read-sachdev} that frustration often induces ground states in which the
translationaly symmetry is broken. In the strong-coupling regime of the
disordered phase of $S=1/2$ systems, the system is made of a collection 
of singlets or plaquettes. This strong coupling regime cannot be described 
starting from independent chains which are gapless. 
Starting from a single chain, small transverse perturbations can yield a 
gap within the TSDMRG. But this gap is often small
and it is difficult to obtain reliable extrapolations. The two-leg
ladder does not present this problem. It does already present a large gap
$\Delta \approx 0.5$ even in absence of frustration. Coupled ladders naturally
evolve toward the 2D N\'eel state as the number of legs increases. Hence,
in principle, disordered and ordered phases could be described within a 
two-leg ladder version of the TSDMRG. This suggests that the two-leg ladder is a more 
natural starting point to describe ground state phases of 2D antiferromagnets 
than the single chain. 
 
  Additional insights into this idea came from my comparative study of
coupled chains with half-integer and integer spins \cite{moukouri-TSDMRG3}. 
In Ref.\cite{moukouri-TSDMRG3}, when starting from single chains, I found 
that although chains with $S=1$ display the Haldane gap, $\Delta \approx 0.4$,
they converge much faster to the N\'eel state than those with $S=1/2$. 
Furthermore, when a frustration induced disordered phase is present, it can be
much more easily found in the case $S=1$. Hence, following the equivalence
between the two-leg ladder and the Haldane spin chain, suggested by the 
Affleck-Kennedy-Lieb-Tasaki construction \cite{aklt}, it
would be better to adopt the two-leg ladder as the building block for 
two-dimensional lattices.

I will now illustrate this idea in the case of the HMCL. Following the
usual notation, the HMCL is given by:

\begin{eqnarray}
  H=J_1 \sum_{<i,j>}{\bf S}_i{\bf S}_j+
J_2 \sum_{[i,j]}{\bf S}_i{\bf S}_j,
\label{hamiltonian}
\end{eqnarray}

\noindent where $<i,j>$ represents nearest-neighbor sites and $[i,j]$
stand for next-nearest neighbors on every other plaquette. $J_1$ is set as
the unit energy. 

 The TSDMRG  with ladders is similar to the method with 
 chains. So I refer the reader to Ref.\cite{moukouri-TSDMRG, 
moukouri-TSDMRG2} for a complete exposition of
the algorithm. Here, I will discuss only briefly the main points of the
algorithm. I start by dividing the 2D lattice into two-leg ladders; the 
Hamiltonian (\ref{hamiltonian}) is written as,

\begin{eqnarray}
  H=\sum_{ladders}H_{ladder}+H_{int},
\label{hamiltonian2}
\end{eqnarray}
 
\noindent where $H_{ladder}$ is the Hamiltonian of a single two-leg ladder,
$H_{int}$ contains the inter-ladder part. In the first step of the
method,  the usual DMRG method is applied to generate a low
energy Hamiltonian of an isolated ladder of  $N_x$ sites keeping $m_1$ states.
Then $m_2$ low-lying states of the superblock states, the corresponding
energies, and all the local spin operators are kept. These energies represent 
the renormalized low energy Hamiltonian of a single ladder. The Hamiltonian
(\ref{hamiltonian2}) is then projected onto the tensor product basis of
independent ladders,

\begin{equation}
\Psi=\prod_{ladders} \Phi_{ladder},
\end{equation}

\noindent where $\Phi_{ladder}$ is an eigenfunction of $H_{0,ladder}$. 
This yield an effective Hamiltonian,

\begin{eqnarray}
  H_{eff}=\sum_{ladders}H_{0,ladder}+{\tilde H}_{int}.
\label{hamiltonian3}
\end{eqnarray}

\begin{figure}
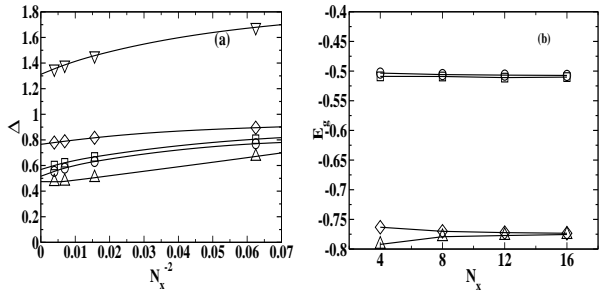

\begin{center}
$\begin{array}{c@{\hspace{0.5in}}c}
         \multicolumn{1}{l} {}\\ [-0.23cm]
\includegraphics[width=1.5 in, height=1.5 in]{gap_lad.eps}
\hspace{0.25cm}
\includegraphics[width=1.5 in, height=1.5 in]{eg.eps}
\end{array}$
\end{center}
\caption{(a) Spin gap of the two-leg ladder for $J_2=0$ (circles), 
$0.5$ (squares), $1$ (diamonds), $1.1$ (triangles up), and $2$ 
(triangles down). (b) Ground-state energies as function of the system size
for a two-leg ladder for $J_2=1$ (circles), $J_2=2$ (diamonds), and for
the 2D lattice for $J_2=1$ (squares), $J_2=2$ (triangle up).}

\vspace{0.5cm}
\label{gaplad-eg}
\end{figure}

 The resulting effective coupled ladder problem which is 1D is studied 
again by the DMRG method in the transverse direction. The TSDMRG is 
variational, as the original DMRG method, the subspace spanned 
by the wavefunctions of the
form $\Psi$ is a subspace of the full Hilbert space of 
Hamiltonian (\ref{hamiltonian}). Its convergence depends on  $m_1$ and $m_2$, 
the error is given by $max(\rho_1,\rho_2)$, where $\rho_1$ and $\rho_2$ 
are the truncation errors in the first and second steps respectively.
$m_2$ fixes the energy band-width $\delta E$. The method is accurate only
when the inter-ladder couplings are small with respect to $\delta E$. 
In the present simulations $\delta E \approx 4$. Since
for the HMCL the inter-ladder and intra-ladder are of the same magnitude,
in principle this approach would be plagued by the same deficiencies the
block RG method. But if the starting point is chosen so that
the essential physics is already contained at the level of the ladder, the
effective strength of the inter-ladder couplings will be small even if the
bare couplings are not. This is particularly the case of models with
frustration in which the competing interactions largely cancel each other
in the strong frustration regime, yielding weakly coupled sub-clusters.

\begin{figure}
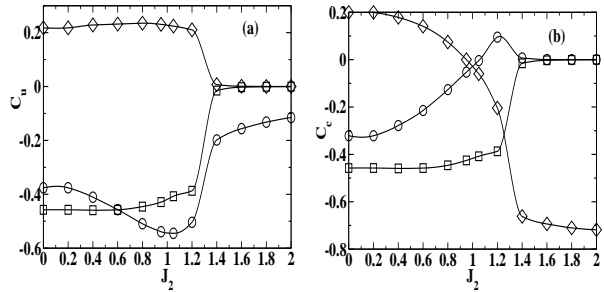

\begin{center}
$\begin{array}{c@{\hspace{0.5in}}c}
         \multicolumn{1}{l} {}\\ [-0.23cm]
\includegraphics[width=1.5 in, height=1.5 in]{src_cb1.eps}
\hspace{0.25cm}
\includegraphics[width=1.5 in, height=1.5 in]{src_cb2.eps}
\end{array}$
\end{center}
\caption{Short-range correlations $C_l$ (circles), $C_r$ (squares), $C_d$ 
(diamonds) for the two-leg ladder for uncrossed (a) and crossed (b) 
plaquettes as function of $J_2$} 
\vspace{0.5cm}
\label{src}
\end{figure}

\begin{figure}
\includegraphics[width=2. in, height=1.5 in]{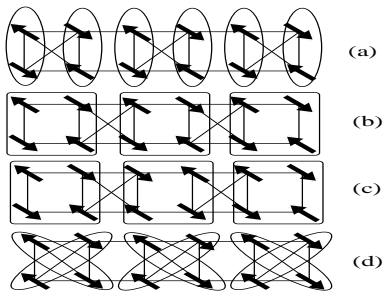}
\caption{ The four phases of the two-leg ladder: rung dimers (a),
plaquette I (b), plaquette II (c), and crossed dimers (d).}
\vspace{0.5cm}
\label{ph}
\end{figure}

The ground state properties of an isolated ladder can readily be obtained.
I keep up to $m_1=96$ and $N_x=16$ and I target spin sectors from
$S_z=0$ to $S_z=\pm 4$ and used open boundary conditions (OBC). 
The maximum error is $\rho_1= 1\times 10^{-4}$. There
is a gap $\Delta$ for all values of $J_2$ investigated between $0$ and
$2$. The finite size behavior of gaps for some typical values of $J_2$ are shown
in Fig.\ref{gaplad-eg}. The case $J_2=0$ reduces to the usual two-leg ladder
which has been widely studied in the literature \cite{acm}. For
 $J_2=0$, $\Delta \approx 0.5$ . As $J_2$ increases, $\Delta$ has a 
non-monotonous behavior.  This suggests a rich structure which is
revealed more clearly by the analysis of the correlation functions.
I computed the following short-range correlation functions: the bond 
strength along a leg $C_{l_{u,c}}=\langle {\bf S}_{i,1}
{\bf S}_{i+1,1} \rangle_{u,c}$ for uncrossed (u) and crossed (c) plaquettes, 
the diagonal correlation 
$C_{d_{u,c}}=\langle {\bf S}_{i,1}{\bf S}_{i+1,2} \rangle_{u,c}$, and
bond strength along the rungs $C_{r}=\langle {\bf S}_{i,1}
{\bf S}_{i,2} \rangle$. Note that I have introduced a second index to
the local spin. These correlations are shown in Fig.\ref{src}. Four
regions can be identified: (i) region I (rung dimers): $0 \alt J_2 \alt 0.6$, 
$C_{l_{u,c}} < 0$, $C_{l_u} \approx C_{l_c}$, $C_{d_{u,c}} > 0$, $C_r <0$,
and $|C_r| > |C_{l_u}|$; the dominant spin-spin correlations are along
the rungs. The ground state properties of the ladder in this 
region are identical to those  of the unfrustrated ladder ($J_2=0$). 
(ii) Region II (plaquettes I): $0.6 \alt J_2 \alt 1$,
$C_{l_{u,c}} < 0$, $|C_{l_u}| > |C_{l_c}|$, $C_{d_{u,c}} > 0$, $C_r <0$,
and $|C_r| < |C_{l_u}|$; the physics is dominated by that of the isotropic 
point. At this point, the ground state is a collection of weakly interacting 
uncrossed plaquettes. Both $C_{d_c}$ and $C_{l_c}$ vanish at $J_2=1$. 
In this region, the local spin configuration is the same on all the 
uncrossed plaquettes
 as shown in Fig.\ref{ph}(b). (iii) region III (plaquettes II): 
$1 \alt J_2 \alt 1.3$, $C_{l_{u}} < 0$, $C_{l_c} > 0$, $C_{d_{u}} > 0$, 
$C_{d_{c}} < 0$, $C_r <0$, and $|C_r| < |C_{l_u}|$; in this region, 
the ground state is again dominated by uncrossed plaquettes. But now the 
local spin configurations on two consecutive uncrossed plaquettes are images
of one another by reflection with respect to a plane passing through the 
middle of the crossed plaquette between them. 
Region IV (crossed dimers): $1.3 \alt J_2 $, $C_{l_{u}} < 0$, 
 $C_{d_{c}} < 0$, $C_{l_c}=C_r =C_{d_{u}} \approx 0$, and 
$|C_r| < |C_{l_u}|$; the ground state is dominated by the crossed dimers on
crossed plaquettes as shown in Fig.\ref{ph}(d). The sketch of the 
spin structure corresponding to each region is summarized in Fig.\ref{ph}.
Since I applied OBC, for a given size, there are two possible 
ground states depending on the plaquette pattern: {\bf (a)} $ucu...ucu$ or 
{\bf (b)} $cuc...cuc$. In region I, the configurations {\bf (a)} and {\bf (b)}
 have nearly the same energy. This is consistent with the fact that the
translational symmetry is not broken. But in Region II and III, {\bf (a)} has
the lowest energy, since it has a larger number of uncrossed plaquettes.
By contrast, in Region IV where dimer order is dominant, it is {\bf (b)}
that has the lowest energy.

The 2D systems are obtained by applying the DMRG on $H_{eff}$ in the 
transverse direction.
I studied systems of size $N_x \times N_y=4\times6,~8\times10,
~12\times14$ and $16\times18$.
I kept up to $m_2=96$ and used OBC. Inter-ladder interactions will have 
very different effects depending on
whether they correspond to a magnetic regime or a disordered regime. 
I will first consider their effects on region II, which includes the 
isotropic point. Recently, there have been a number of studies which strongly
suggest that the physics of the 2D systems is identical to that displayed
by the two-leg ladder. In other words, the ground state  is essentially made of weakly
interacting plaquettes. If this is the case, it means that the inter-ladder 
interactions will not strongly modify the ground state wave function  
of decoupled ladders.  Fig.\ref{gaplad-eg} shows that the 
ground state energy and $\Delta$ remain very close to that of an isolated
plaquette. Thus in the vicinity of $J_2=1$, inter-ladder interactions do
not strongly renormalize the properties of an isolated ladder which
themselves are close to those of an isolated plaquette. The extrapolated
gap is found to be $\Delta =0.67 J_1$ which is in good agreement with
the prediction from exact diagonalization \cite{fouet}. The same conclusion
is seen in Fig.\ref{gaplad-eg} for region IV where the crossed-dimer ground state  
found for the ladder is also the ground state of the 2D lattice. In both
cases, the wave function made of the tensor product of the wave function
of single two-leg ladders is a good variational wave function for the 2D
system. In each case, the ground state energy of the 2D system remains
very close to that of individual plaquettes or crossed dimers. This
can be explained as follows: when ladders are brought together to build 
the 2D lattice, the dominant local correlations are $C_{l_u}$ 
in region II and $C_{d_c}$ in region IV; during this process,
magnetic energy cannot  efficiently be gained. For region II, this 
is because the two 
neighboring plaquettes of an uncrossed plaquette in the direction of the 
rungs involve frustrated bonds. Hence the system prefers the original
configuration to avoid increasing its energy. For region IV, $C_r$ is 
very small. The system cannot increase it when the ladders are coupled, because
the spins are already involved in strong diagonal dimers. There is,
however, the possibility to gain magnetic energy by forming N\'eel order 
along the direction of the diagonal bonds ($J_2$  direction) as suggested 
in Ref.\cite{sfb}. This is unlikely, however,  because once such
a phase is reached, I do not see how the system could go to crossed dimers
at larger $J_2$. The action of $J_1$ which act as frustration in this regime
decreases as $J_2$ increases. Hence once this hypothetical  N\'eel phase along 
the $J_2$ bonds is reached, there is no obvious mechanism that could destroy 
it as $J_2$ increases to yield the crossed-dimer phase as suggested in 
Ref.\cite{sfb}. Such a N\'eel phase would be favored only 
when $J_2 \gg J_1$. I made rough calculations with $J_2=4,8$ and I found 
that the system remains in the crossed-dimer phase. The situation 
is apparently identical to the $J_1-J_2$ chain where the independent 
chains regime is only reached in the infinite $J_2$ limit.

\begin{figure}
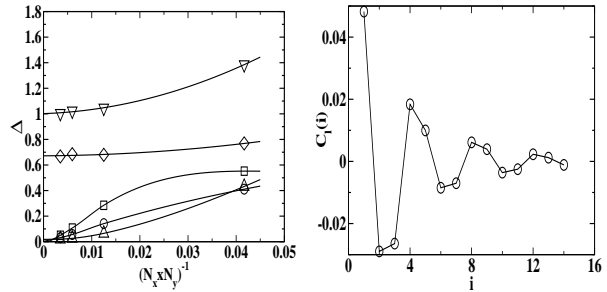

\begin{center}
$\begin{array}{c@{\hspace{0.5in}}c}
         \multicolumn{1}{l} {}\\ [-0.23cm]
\includegraphics[width=1.5 in, height=1.5 in]{gap_2D.eps}
\hspace{0.25cm}
\includegraphics[width=1.5 in, height=1.5 in]{corlgp1.25l16.eps}
\end{array}$
\end{center}
\caption{ (a) 2D gaps as function of the system size $J_2=0$ (circles), 
$0.5$ (squares), $1$ (diamonds), $1.1$ (triangles up), and 
$2$ (triangles down). (b) Correlation function along the legs as function
of the distance for $J_2=1.25$.}
\vspace{0.5cm}
\label{gap-lrc}
\end{figure}

\begin{figure}
\includegraphics[width=3. in, height=1. in]{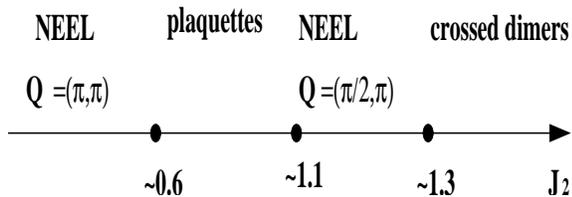}
\caption{ Ground-state phases of the 2D checkerboard model as function 
of $J_2$. Note that the phase boundaries are rough estimates taken from the 
phases of the two-leg ladder. }
\vspace{0.5cm}
\label{pd-2D}
\end{figure}

The situation is very different for regions I and III. In region I,
the dominant local correlation is $C_r$; when the ladders are brought
together, magnetic energy can be gained by
an antiferromagnetic arrangement along the rungs. This enhances the
local antiferromagnetic order which exists along the legs and ultimately leads to
a N\'eel order with $Q=(\pi,\pi)$. This is seen in the vanishing of
the spin gap for $J_2=0$ and $J_2=0.5$ shown in Fig.\ref{gap-lrc}(a).
This is in agreement with results for $J_2=0$ from quantum Monte Carlo
(QMC) simulations \cite{sandvik}
and large $S$ analysis \cite{canals}. I find that
the TSDMRG ground state energy $-0.6011$ at $J_2=0.$ is not in very good 
agreement with the QMC result $-0.6699$ of Ref.\cite{sandvik}. 
Despite this discrepancy, the TSDMRG is nevertheless
able to reproduce the low-energy behavior of the ordered phase. This is not
in fact surprising. In the Resonating valence bond picture, the N\'eel 
state and its low energy excitations can be written as linear combination
of a tensor product of dimers. The TSDMRG variational solution of 
Hamiltonian(\ref{hamiltonian})  which is a linear combination of 
the wave functions $\Psi$ has exactly this form.  
A similar analysis also applies for region III. $C_{l_u}$ is dominant 
in region II. But as seen in Fig.\ref{src}, $C_{l_u}$ has a minimum at 
$J_2=1$ and then increases. It becomes very close to $C_r$ when $J_2$ 
enters region III.
Hence magnetic energy can be gained again through the rungs. Since
the structure along the legs is not modified from Fig.\ref{ph}(d), 
the resulting wave vector will be $Q=({\pi/2,\pi})$. This is seen in 
Fig.\ref{gap-lrc} in the behavior of the spin-spin correlation function
$C_l(i)$ along the legs. $C_l(i)$ displays a period of $4$. 
 The correlations between the rungs (not shown) oscillate with $q_y=\pi$.  
Fig.\ref{pd-2D} presents a sketch of the different ground state phases 
of the HMCL as function of $J_2$. I note that in Ref.\cite{sfb}, a 
very similar phase diagram was suggested; the only difference with 
the TSDMRG phase diagram is the wave vector of the N\'eel phase
between the plaquette and crossed-dimer phases.

In summary, I have shown  that the TSDMRG method can
reliably be used to study the disordered phases with short correlation
lengths of isotropic 2D models. In these phases, the system is
a collection of dimers or plaquettes. This makes the two-leg
ladder a very good  starting point for a variational calculation. I showed
that the basic physics of 2D systems could already be read through
short-range correlations of the two-leg ladders. 
This variational calculation is able to predict reasonably 
magnetically ordered phases as well. In this work, I did not 
discuss the question of low-lying singlet excitations within the gap. 
Targeting them will lead
to large truncation errors and the calculations will become impractical.
These excitations are naturally truncated out when they are not needed
to form a target state. Finally, The same method could be applied
to  the Sutherland-Shastry, $J_1-J_2$ or the Kagom\'e models in 2D.
It could also be applied to the pyrochlore lattice,
provided that the Hamiltonian could be written in some form
involving 1D subsystems with a large gap.

\begin{acknowledgments}
 This work started during a visit at the Weizmann Institute. The author
thanks E. Altman for hospitality. This work was supported by the NSF 
Grant No. DMR-0426775.
\end{acknowledgments}

\end{document}